\documentclass[a4paper,fleqn,usenatbib]{mnras}

\usepackage{newtxtext,newtxmath}

\usepackage[T1]{fontenc}
\usepackage{ae,aecompl}

\usepackage{graphicx}	
\usepackage{amsmath}	
\usepackage{epsfig}
\usepackage{animate}

\title[3D Optical Diagnostics]{Constraining Photoionization Models With a Reprojected Optical Diagnostic Diagram}

\author[X. Ji \& R. Yan]{Xihan Ji$^{1}$\thanks{Contact e-mail: \href{mailto:xji243@uky.edu}{xji243@uky.edu}}, Renbin Yan$^{1}$\thanks{Contact e-mail: \href{mailto:rya225@g.uky.edu}{rya225@g.uky.edu}}
\\
$^{1}$Department of Physics and Astronomy, University of Kentucky, 505 Rose Street, Lexington, KY 40506, USA}

\date{Accepted XXX. Received YYY; in original form ZZZ}

\pubyear{2020}

\begin{document}
\label{firstpage}
\pagerange{\pageref{firstpage}--\pageref{lastpage}}
\maketitle

\begin{abstract}
Optical diagnostic diagrams are powerful tools to separate different ionizing sources in galaxies. However, the model-constraining power of the most widely-used diagrams is very limited and challenging to visualize. In addition, there have always been classification inconsistencies between diagrams based on different line ratios, and ambiguities between regions purely ionized by active galactic nuclei (AGNs) and composite regions. We present a simple reprojection of the 3D line ratio space composed of [N~II]$\lambda 6583$/H$\alpha$, [S~II]$\lambda \lambda$6716, 6731/H$\alpha$, and [O~III]$\lambda$ 5007/H$\beta$, which reveals its model-constraining power and removes the ambiguity for the true composite objects. It highlights the discrepancy between many theoretical models and the data loci. With this reprojection, we can put strong constraints on the photoionization models and the secondary nitrogen abundance prescription. We find that a single nitrogen prescription cannot fit both the star-forming locus and AGN locus simultaneously, with the latter requiring higher N/O ratios. The true composite regions stand separately from both models. We can compute the fractional AGN contributions for the composite regions, and define demarcations with specific upper limits on contamination from AGN or star formation. When the discrepancy about nitrogen prescriptions gets resolved in the future, it would also be possible to make robust metallicity measurements for composite regions and AGNs.

\end{abstract}




\section{Introduction}

The widely used optical diagnostic diagrams (hereafter BPT diagrams) originally proposed by \cite{1981PASP...93....5B} and refined by \cite{1987ApJS...63..295V} are very useful for distinguishing among different ionization mechanisms in galaxies. The merits of these diagnostics are that only ratios of strong optical emission lines are involved (i.e. [S~II]$\lambda \lambda$6716, 6731/H$\alpha$, [N~II]$\lambda$6583/H$\alpha$, [O~I]$\lambda$6300/H$\alpha$, and [O~III]$\lambda$5007/H$\beta$), and they are insensitive to dust extinction. Various demarcations have been proposed based on these diagrams over the years, which add to our understanding of the variety and complexity of the ionization mechanisms in the universe \citep{1987ApJS...63..295V, 1997ApJS..112..315H, 2001ApJ...556..121K, 2003MNRAS.346.1055K, 2006MNRAS.372..961K, 2008ARA&A..46..475H}. Utilizing sophisticated photoionization codes like {\tt CLOUDY} and {\tt MAPPINGS}, people are able to match the distribution of the observed data in the BPT diagrams with the model predictions, and put constraints on the properties of ionized regions and/or sources \citep[e.g.][]{2000ApJ...542..224D, 2002ApJS..142...35K, 2004ApJS..153...75G, 2013ApJS..208...10D}.

However, there are several limitations associated with the usage of the BPT diagrams. First of all, there is often ambiguity between the classifications of different BPT diagrams. Spectra classified as AGN in one diagram could be classified as star-forming (SF) in another diagram, and vice versa. These ambiguous cases are tricky to deal with and are often excluded in data analyses, leading to biased or incomplete physical interpretations. In addition, photoionization predictions with significantly different model parameters (densities, abundance patterns, ionizing spectra, etc.) are able to cover the data space equally well \citep[e.g.][]{2019ApJ...878....2D}. The lack of a merit function that requires consistency across multiple diagrams makes it difficult to evaluate the goodness of the fit. Partly this is because of the difficulty of visualizing model consistency across different diagnostic diagrams. Finally, apart from the degeneracy in the model parameters, there is also difficulty in the interpretation of the zone in-between the SF locus and AGN region in the BPT diagrams, which is commonly named as `composite', `transition', or `ambiguous' region. The line ratios there could either be a result of mixing between different ionized clouds, e.g. SF and Seyfert, or SF and low-ionization nuclear emission-line regions (LINERS), due to insufficient spatial resolution, or it could be caused by the intrinsic variations in the physical parameters of Seyfert narrow-line regions or LINERs. It is hard to break this degeneracy with the traditional BPT diagrams. There are numerous efforts trying to decompose points in this zone into combinations of AGN and SF \citep[e.g.][]{2006MNRAS.372..961K}. However, the semi-arbitrary choices of the starting points in the AGN and SF zones in the BPT diagrams make those results unconvincing. In addition, this degeneracy makes it difficult to constrain metallicities or ionization parameters for data points in this zone and they are hardly attempted.

The above issues could be largely resolved if one requires the consistency of the model predictions on more than two line ratios. A simple solution would be to compare models with data in a multi-dimensional line-ratio space. The traditional BPT diagrams can be viewed as specific projections of this higher-dimensional space. However, these straight-forward projections are poorly positioned in revealing the discrepancy or consistency between models and data. A carefully chosen new projection from a more advantageous angle would be able to show those discrepancies more clearly. The angle of such a projection should be determined by the intrinsic shapes and orientations of the model surfaces embedded in higher dimensions. In this paper, we utilize the three traditional and most commonly available line ratios, [N~II]$\lambda 6583$/H$\alpha$, [S~II]$\lambda \lambda$6716, 6731/H$\alpha$, and [O~III]$\lambda$ 5007/H$\beta$, to construct a new set of re-projected optical diagnostic diagrams, which have the power to overcome all of the issues mentioned above.

The concept of requiring consistency across multiple line ratios have been applied in some cases, especially for the derivation of metallicities. \cite{2004ApJ...613..898T}, for instance, used six strong optical emission lines and calculated the likelihood distribution of metallicity for their sample galaxies by comparing the measured line intensities with a large number of models. Similarly, by adopting the Bayesian inference, \cite{2015ApJ...798...99B} developed the code {\tt IZI} to consistently infer the metallicity and ionization parameter of H~II regions given a set of observed strong emission lines and star-forming templates. In principle, these methods can also be applied to AGN narrow-line regions so long as the corresponding models are computed. However, without a viable visualization of the consistency between model and data in the higher-dimensional space, models that are systematically offset from the data in the higher dimensions can still be used in these methods without users realizing it. 
In addition, these methods all rely on prior classifications into AGN and SF according to BPT diagrams, which would suffer the same issues mentioned above. 
Last but not least, all these methods would have difficulties when dealing with the `composite regions'.

\cite{2014ApJ...793..127V}, on the other hand, directly looked into the distribution of galaxies in the three-dimensional line-ratio space, and constructed a number of three-dimensional diagnostic diagrams using different combinations of optical emission line ratios. By incorporating photoionization models for SF regions, they defined a series of two-dimensional diagnostic diagrams projected from three dimensions. These diagrams are capable of showing the metallicity sequence of SF regions clearly, and their classification of different ionized regions is in good agreement with that based on the original [N II] BPT diagram. In this work, however, we take a further step by considering the photoionization models for AGN regions at the same time. Our preferred projection is defined by making not only the SF model surface, but also the AGN model surface compact. This definition puts a stronger constraint on the resulting viewing angles in three dimensions, and facilitates analyses on the composite regions. In addition, we carry out a detail comparison among models with different input parameters in the derived projection, in order to make a self-consistent definition of theoretical demarcations for SF and AGN regions.

In summary, while various attempts have been made in deriving the properties of ionized regions in high dimensions using photoionization models, seldom are the model parameters and assumptions carefully examined. Therefore, it is important to first verify the consistency between models and data in the higher-dimensional space, which would then allow much more robust constraints on metallicity and ionization parameters.
Our solution, a set of reprojected diagnostic diagrams obtained through rearranging the line ratios used in the original optical diagrams, not only enables strong constraints on the model assumptions, but also has the potential to make metallicity constraints much more robust and applicable to the composite regions. 

In \S\ref{sec:model_data}, we introduce the theoretical models and observational data we use. We show the comparison between this new diagram and the original BPT diagrams, and describe the procedure to construct this new diagram in \S\ref{sec:new_proj}. In \S\ref{sec:application}, we discuss the implications and applications of the new diagram. Discussions about the robustness of our analyses, including the impact of sample selection on data distribution and the effect of time evolution on our photoionization models, are given in \S\ref{sec:discuss}. We summarize our conclusions in \S\ref{sec:conclude}. Throughout this paper, 
all wavelengths are given in air. For the three frequently used line ratios in this paper, i.e. [N~II] $\lambda 6583$ / H$\alpha$, [S~II] $\lambda \lambda$ 6716, 6731 / H$\alpha$ and [O~III] $\lambda$ 5007 / H$\beta$, we denote their decadic logarithms as N2, S2, and R3, respectively.

\section{Models and data} \label{sec:model_data}

\begin{table*}
	\centering
	\caption{Photoionization model sets}
	\label{tab:models}
	\begin{tabular}{l c}
	    \hline
		\hline
		Parameter & Values \\
		\hline
		\multicolumn{2}{|c|}{SF models} \\
		\hline
		q & $-4.0$, $-3.5$, $-3.0$, $-2.5$, $-2.0$\\
		$[{\rm O/H}]$ & $-1.3$, $-0.7$, $-0.4$, 0.0, 0.3, 0.5\\
	    $\log (n{\rm_H} / {\rm cm}^{-3})$ & 1.15\\
		Ionizing SED & {\tt Starburst99} models with $\log (Z/Z_\odot)$ = $-1.3$, $-0.7$, $-0.4$, 0.0, 0.3 \\
		 & Continuous SFH for 4 Myr\\
		Nitrogen prescription & Dopita13 prescription\\
		\hline
		\multicolumn{2}{|c|}{AGN models} \\
		\hline
		q & $-4.0$, $-3.5$, $-3.0$, $-2.5$, $-2.0$\\
		$[{\rm O/H}]$ & $-0.75$, $-0.5$, $-0.25$, 0.0, 0.25, 0.5, 0.75\\
		$\log (n{\rm_H} / {\rm cm}^{-3})$ & 2.0\\
		Ionizing SED & power-law SED with the power-law index $\alpha$ = $-1.4$ (fiducial), $-1.7$ (preferred by data) \\
		Nitrogen prescription & Groves04 prescription\\
		\hline
	\end{tabular}
\end{table*}

The new optical diagnostic diagram we propose in this paper is determined by the overall shapes and orientations of the SF and AGN model grids in the line-ratio space, as this is a projection that compactify both models simultaneously. A summary of the major parameters of the models is shown in Table~\ref{tab:models}. Note that this table only includes our fiducial and best-fit models, and we will compare them to other models with different densities, ionizing SEDs, or nitrogen prescriptions in \S\ref{subsec:mparam}. The following is a description of the models and the data we use.

These models are generated using the photoionization code {\tt CLOUDY} (\citealp{2017RMxAA..53..385F}). The input SEDs for the SF models are computed using the code {\tt Starburst99} \citep{1999ApJS..123....3L}. We assume a Kroupa initial mass function (IMF), and a continuous star formation history over 4 Myr. SEDs with different stellar metallicities are computed, with  $Z/Z_{\odot}$ = 0.05, 0.2, 0.4, 1, and 2, respectively. The ionized cloud is set to have a hydrogen density of 14 cm$^{-3}$ (at the ionizing face), which is the median value we obtain using the {\tt temden} routine in {\tt PyRAF} (\citealp{1987JRASC..81..195D, 1995PASP..107..896S}) for H~II regions in MaNGA, assuming an electron temperature of 10$^4$ K. The gas pressure is assumed to be constant throughout the cloud. We include dust grains with typical ISM abundance, which will also scale with the metallicity of the cloud. Metal depletion onto the dust grains is computed using the values given by \cite{1987ASSL..134..533J} and \cite{1986ARA&A..24..499C}. Finally, the cosmic ray background of the local Universe is included.

To set up a grid, we vary the ionization parameter and the gas phase metallicity of the cloud. The ionization parameter is defined as  $ q\equiv \log U \equiv \log (\rm \Phi _{ion}/n_{\rm H}c)$, where $\rm \Phi _{ion}$ is the flux of ionizing photons at the illuminated surface of the cloud, and n$\rm_H$ is the volume density of hydrogen. The range we adopt for this parameter, $q$, is from $-$ 4.0 to $-$ 2.0. In the meanwhile, we vary the logarithmic gas-phase metallicity [O/H] $\equiv \log ( ({\rm O/H})/ ({\rm O/H})_\odot)$ from $-$ 1.3 to 0.5 (corresponding to 0.05 to 3.16 in linear space). For the SF models, we adopt a consistent stellar metallicity for the input SED as the gas metallicity when possible. Since our input SEDs only cover stellar metallicity up to $Z/Z_{\odot}$ of 2, for all models with gas metallicities greater than this value, we use the SED with the highest metallicity. All elements except for helium, carbon, and nitrogen directly scale with the oxygen abundance. The solar abundance we use is taken from \cite{2010Ap&SS.328..179G}, with 12 + log (O / H)$_{\odot}$ = 8.69. The helium abundance is described by a primary production formula (\citealp{2002ApJ...572..753D}), which gives the number density ratio of helium to hydrogen:
\begin{equation}
    \text{He / H} = 0.0737 + 0.024\cdot \text{(O/H) / (O/H)}_\odot .
\end{equation}

For carbon and nitrogen, contribution from the secondary production is important. 
How to quantitatively account for the yield of the secondary elements remains debated among literature. And there is evidence that the abundance of these elements could have dependence on the properties of galaxies, like the total stellar mass and the star formation efficiency \citep[e.g.][]{2017MNRAS.469..151B, 2020ApJ...890L...3S}. In this paper, we use the nitrogen prescription adopted by \cite{2013ApJS..208...10D} for our SF models. \cite{2013ApJS..208...10D} used a compilation of observed H~II regions from \cite{1998AJ....116.2805V}. We refit this data using a simple quadratic function:
\begin{equation}
   \text{N / O} = 0.0096 + 72\cdot \text{O / H} + 1.46\times 10^4 \cdot \text{(O / H)}^2 .
   \label{eq_dop}
\end{equation}
When we adopt this Nitrogen prescription,  the carbon abundance is set to be always 1.03 dex larger than the nitrogen abundance, to be consistent with \cite{2013ApJS..208...10D}. 

For the AGN models, however, we switch to the following nitrogen prescription described by \cite{2004ApJS..153....9G} to fit the line ratios of AGN host galaxies.
\begin{equation}
    \rm N/O = 10^{-1.6} + 10^{2.33 + log(O/H)} .
    \label{eq_gro}
\end{equation}
\cite{2004ApJS..153....9G} obtained this prescription by fitting a set of H~II regions and nuclear starburst galaxies from \cite{2002A&A...389..106M} and \cite{2003ApJ...591..801K}. When adopting this prescription, we set the carbon abundance to be 0.6 dex larger than the nitrogen abundance, to be consistent with the solar abundance. We will detail in \S\ref{subsec:mparam} why we decide to apply different prescriptions to different photoionization models.

The form of the ionizing SED we choose for the AGN models is a broken power-law function adopted by \cite{2004ApJS..153....9G}. Our starting model assumes that $d \log F_{\nu} / d \log(\nu) = -1.4$, from $h\nu = 1$ Ryd to approximately 100 Ryd. This simple SED fits the observational data reasonably well in the original BPT diagrams. \cite{Ji2020} showed that this model provides a nice match to the upper boundary of the AGN region shown by their sample in the [SII] BPT diagram. Later in \S\ref{subsec:mparam}, we will use the SED with a power-law index of $-1.7$ in this intermediate energy range, as photoionization models with this softer SED match the majority of the AGN-ionized clouds better in the new diagram. The hydrogen density is set to be 100 cm$^{-3}$, which is the median value we found for the narrow line regions (NLRs) in MaNGA \citep{Ji2020}. The ionization parameter for the ionized cloud varies from $-$ 4.0 to $-$ 2.0 and the logarithm of the metallicity varies from $-$ 0.75 to 0.75.

The observational data we use are obtained by MaNGA \citep{2015ApJ...798....7B, 2016AJ....152..197Y} and released as part of the 15th data release (DR15) of Sloan Digital Sky Survey (SDSS), which is identical to the 7th internal MaNGA Product Launch (MPL-7). This data set includes spatially-resolved spectroscopic data of 4621 unique galaxies. MaNGA is part of SDSS-IV \citep{2017AJ....154...28B}, and is the largest integral field spectroscopy (IFS) survey of galaxies to date. The observation is done using the 2.5 m Sloan Telescope \citep{2006AJ....131.2332G}. MaNGA utilizes the BOSS spectrographs \citep{2013AJ....146...32S} and a number of  fiber-bundle integral field units \citep{2015AJ....149...77D} to provide spectra with a wavelength coverage from 3622\AA ~to 10,354\AA, and a median spectral resolution of $R\sim 2000$. 
The data are reduced by the MaNGA Data Reduction Pipeline \citep{2016AJ....152...83L} which produces datacubes with each spatial pixel (spaxel) being $0\arcsec.5\times 0''.5$. The relative flux calibration is accurate to 1.7\% between H$\alpha$ and H$\beta$ \citep{2016AJ....151....8Y}. The median FWHM of MaNGA's PSF is close to $2''.5$ \citep{2015AJ....150...19L}, which corresponds to a physical scale of $\sim 1.5$ kpc at a typical redshift of 0.03 \citep{2017AJ....154...86W}. The emission line measurements we present in this paper are performed by the MaNGA data analysis pipeline \cite[DAP;][]{2019AJ....158..231W, 2019AJ....158..160B}, which uses the code {\tt pPXF} \citep{2004PASP..116..138C, 2017MNRAS.466..798C} at its core.


Our sample consists of spaxels from the central regions of MaNGA galaxies, with $r / R_{\rm e} < 0.3$, where $r$ is the inclination-corrected galactic-centric distance of a given spaxel and $R\rm_e$ is the elliptical Petrosian effective semi-major axis measured in the SDSS $r$-band of a given galaxy. We require that the signal-to-noise ratios (S/N) of the H$\alpha$, H$\beta$, [O~III] $\lambda$ 5007, [NII] $\lambda$ 6583, and [SII] $\lambda \lambda$ 6716, 6731 emission lines are all greater than 3. There is no cut in the ionization properties, so this sample include spaxels of SF-photoionized regions, AGN-photoionized regions, low-ionization nuclear emission-line regions (LINERs) photoionized by AGNs, and low-ionization emission-line regions (LIERs) not photoionized by AGNs. The latter two categories are indistinguishable by strong-line ratios alone and we will group them as LI(N)ERs. We will discuss these LI(N)ER spaxels and their identification in \S\ref{subsec:liner}.

\section{New projection for the optical diagnostic diagram}
\label{sec:new_proj}

\begin{figure*}
    \animategraphics[autoplay,loop,width=0.33\textwidth]{3}{Section3/animation/rotation_}{0}{35}
    \includegraphics[width=0.33\textwidth]{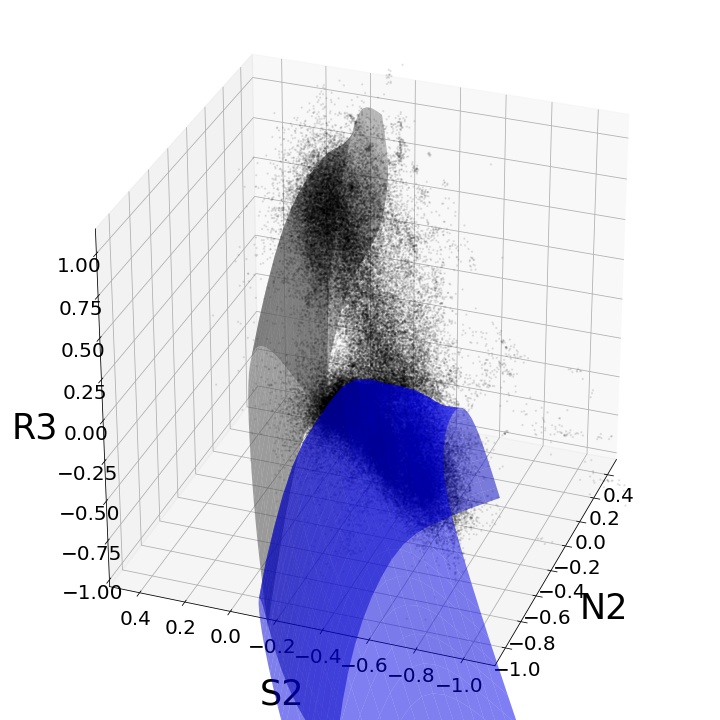}
    \includegraphics[width=0.33\textwidth]{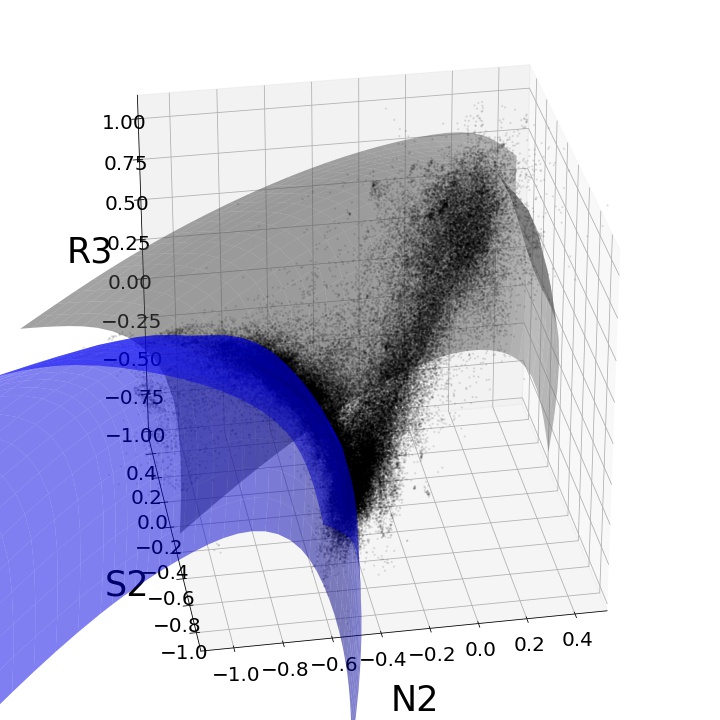}
    \caption{Left: a 3D BPT diagram viewed at an elevation angle of 30 degrees. A sample of MaNGA central spaxels with $r/R_{\rm e} < 0.3$ are shown as black points. The SF and AGN models are plotted as the blue and the grey surfaces, respectively. Click on this image in a PDF viewer to make it stop rotating. Middle and Right: two still images from the animation, illustrating the gap between two model surfaces and how the data points are distributed relative to them.}
    \label{fig:3d_BPT}
\end{figure*}

\begin{figure*}
	\centerline{\includegraphics[width=1.1\textwidth]{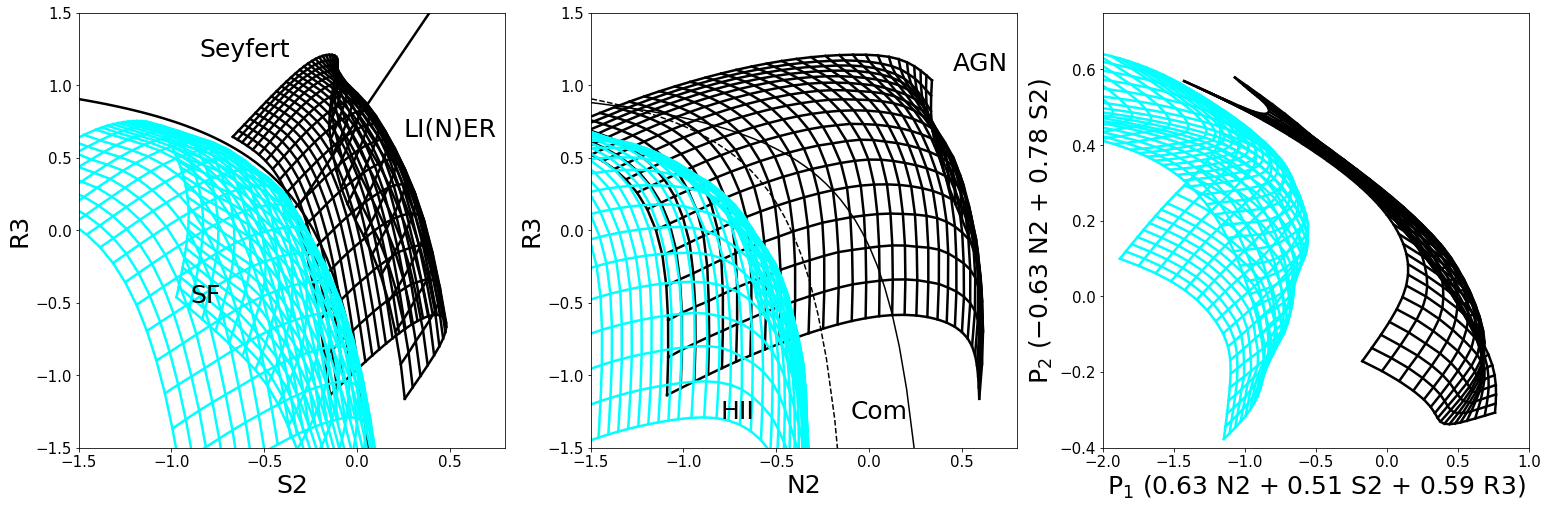}}
	\caption{Comparison of photoionization models in three optical diagnostic diagrams. AGN model grids (black) and SF model grids (cyan) displayed in the diagrams have been smoothed through interpolation. Left panel: the [S~II]-based BPT diagram; middle panel: the [N~II]-based BPT diagram; right panel: the new diagram proposed by this paper.}
    \label{fig:3bpt_com}
\end{figure*}

\subsection{Projections from the 3D line-ratio space}

In Fig.~\ref{fig:3d_BPT}, we plot our photoionization models for SF and AGN together with our sample in an animated three-dimensional line-ratio space spanned by N2, S2, and R3. Each photoionization model manifests itself as a 2D continuous surface embedded in the 3D space, and each point on the model surface corresponds to a unique set of metallicity and ionization parameter. It is clear that there is a gap between the AGN model surface and the SF model surface, indicating that there is no degeneracy between these two ionization models, at least in the part of parameter space we care about. Line ratios of pure SF or pure AGN have to be located on or near these two disjoint model surfaces. Some data points are indeed there. However, some of the data points lie in the gap, separate from both model surfaces. They actually form a bridge connecting the two surfaces. The most natural explanation is that they are indeed `composite' objects resulting from a combination of SF and AGN due to unresolved observations.

In Fig.~\ref{fig:3bpt_com}, we show a few 2D projections of this 3D line-ratio space. Two of the projections correspond to the traditional [S~II]-based and [N~II]-based BPT diagrams, while the third one is a new projection we propose. The axes of the new projected 2D diagram (wihch we denote as $P_1$ and $P_2$) are made of linear combinations of the three line ratios from the two original diagrams, thus having the same virtue of using only strong lines and being insensitive to extinction. The advantage of the new diagram compared to the original ones is that it maximizes the separation between the SF model grids and the AGN model grids. In essence, we are choosing the projection that views both model surfaces roughly edge-on for the parts of parameter space that matter. The gap between them would become even emptier, if one further removes some models that are not found in the data, judging with the 3rd dimension that is perpendicular to the two dimensions shown here. We will illustrate this below.

Furthermore, with the model surfaces appearing edge-on, it becomes easier to directly compare them with the observational data, in a way that maintains consistency across three line ratios. It not only enables us to better constrain the input parameters of the models, but also provides a way to investigate the contribution of different ionization sources to the observed line ratios. We will discuss these applications in \S\ref{sec:application}.

\subsection{Derivation of the axes of the new projection}

\begin{figure*}
	\includegraphics[width=0.9\textwidth]{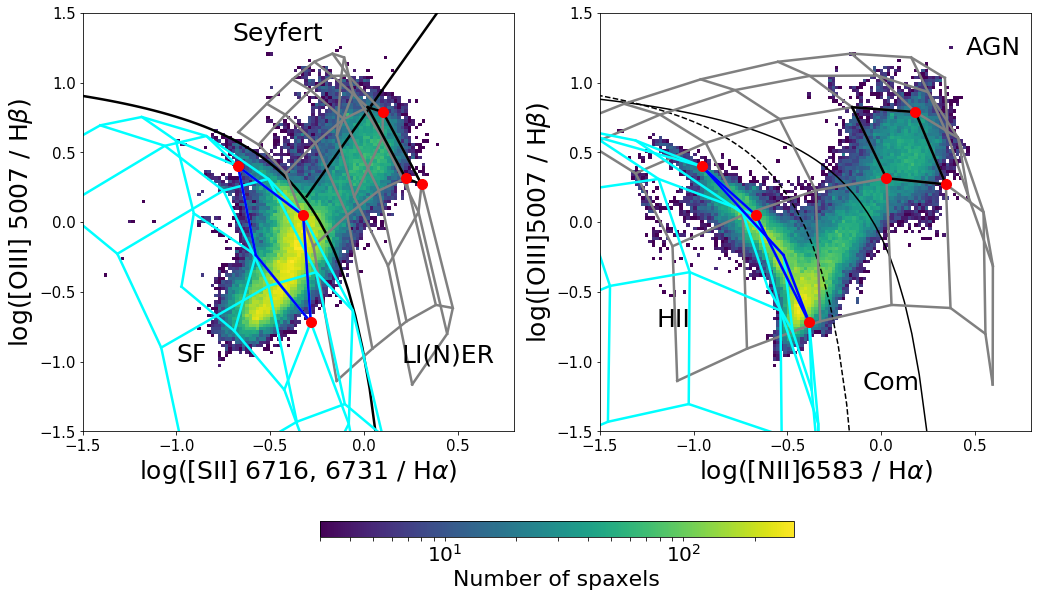}
	\caption{Photoionzation models and observational data viewed in the [S~II]- and [N~II]-based BPT diagrams. The solid black curves are \citet{2001ApJ...556..121K} extreme star-burst lines. The dashed black curve in the right panel is the \citet{2003MNRAS.346.1055K} demarcation that separates H~II regions from AGNs. The solid straight black line in the left panel is the \citet{2006MNRAS.372..961K} line which separate Seyferts from LI(N)ERs.  The models used here are described in \S\ref{sec:model_data}. The parts of the model grids that are highlighted in black (blue) is the patch we select to represent the AGN (SF) model surface. Red points are chosen to construct planes in the 3D diagram to find the angle from which both model surfaces are viewed nearly `edge-on'.}
    \label{fig:patches}
\end{figure*}

Fig.~\ref{fig:patches} shows the data distribution and our models in the traditional BPT diagrams. 
From the 3D perspective, we could view the [N~II] and [S~II] diagrams as two different projections. The AGN model in the [S~II] projection is closer to being `edge-on' while it is nearly `face-on' in the [N~II] one. This is why the data distribution in the [S~II] diagram is more useful for constraining the AGN SED \citep{Ji2020}.
By contrast, the [N~II] projection works better to isolate pure H~II regions, because the SF surface becomes more compact at high metallicity in this diagram. It is natural to imagine that there is a projection that make both surfaces edge-on, which would be more ideal to reveal the gap between model surfaces and to separate regions with different ionization mechanisms.

Here we show how to obtain the ideal projection by consecutive rotations in the 3D line-ratio space. To begin with, we note that for a small enough region on the model grid, the curvature of the model surface is negligible. If we have two planes with exactly zero curvatures, the procedure to find their edge-on projection can be summarized as follows:
\renewcommand{\labelenumi}{\arabic{enumi}}
\begin{enumerate}
    \item Find where the two planes intersect. It will be a straight line in the 3D space;
    \item Find a plane which is perpendicular to this intersection line;
    \item Find the two axes of this plane. We note that in this step, one has the freedom to change these axes under a rotation about the intersection line. An easy way to find a set of axes is to perform two consecutive rotations of the original axes. The direction of the intersection line determines a polar angle $\theta$ and an azimuthal angle $\phi$. If we start with N2, S2, and R3 as $x$, $y$, and $z$ axes, a first rotation about the $z$ axis by an angle $\phi$ and a second rotation about the new $y$ axis by an angle $\theta$ would directly lead us to the plane (which is the final $P_1-P_2$ plane).
\end{enumerate}
Going back to our model surfaces, they are curved surfaces in the 3D diagram. But since we only care about the part of parameter space where our sample is located, we can pick out certain patches of the model surfaces, which will be small enough to be approximated by planes. In Fig.~\ref{fig:patches}, we choose one quadrangle patch from each photoionization model grid. The patch of SF grid covers the part of the parameter space satisfying 0.0 < [O/H] < 0.3 and $-2.5$ < q < $-3.0$, and the patch of the AGN grid covers the parameter space satisfying 0.0 < [O/H] < 0.25 and $-3.0$ < q < $-3.5$. These parts of models are chosen as they represent the line ratios of the the most densely-populated areas by the data in the [N~II] and [S~II] BPT diagrams. The curvature of each patch is relatively small, and we can construct two planes by selecting three points from each of the quadrangle. Using these two planes, we are able to find the desired final plane of projection.

\begin{figure}
	\includegraphics[width=0.5\textwidth]{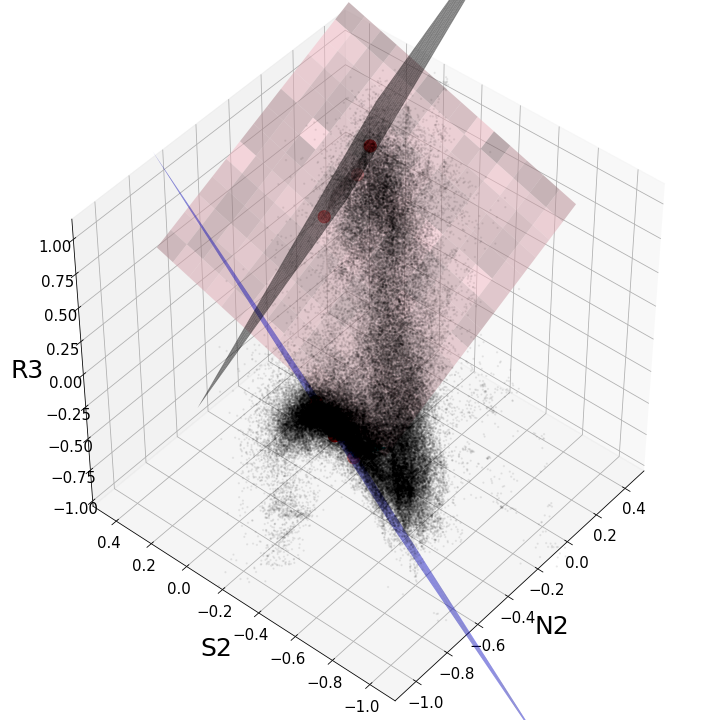}
	\caption{The final plane (pink) constructed from the two model patches (in grey and cyan, respectively). The black (blue) plane is the representative plane for the AGN (SF) model constructed through the red points, and is perpendicular to the final plane. The black data points are our sample spaxels.}
    \label{fig:newr_3d}
\end{figure}

Fig.~\ref{fig:newr_3d} shows how we construct the final plane in the 3D line-ratio space. The polar angle $\theta$ (from the R3 axis) is 36 degree, and the azimuthal angle $\phi$ (defined as a counter-clockwise angle from the positive direction of the N2 axis) is 39 degree. 

\begin{figure*}
	\includegraphics[width=0.95\textwidth]{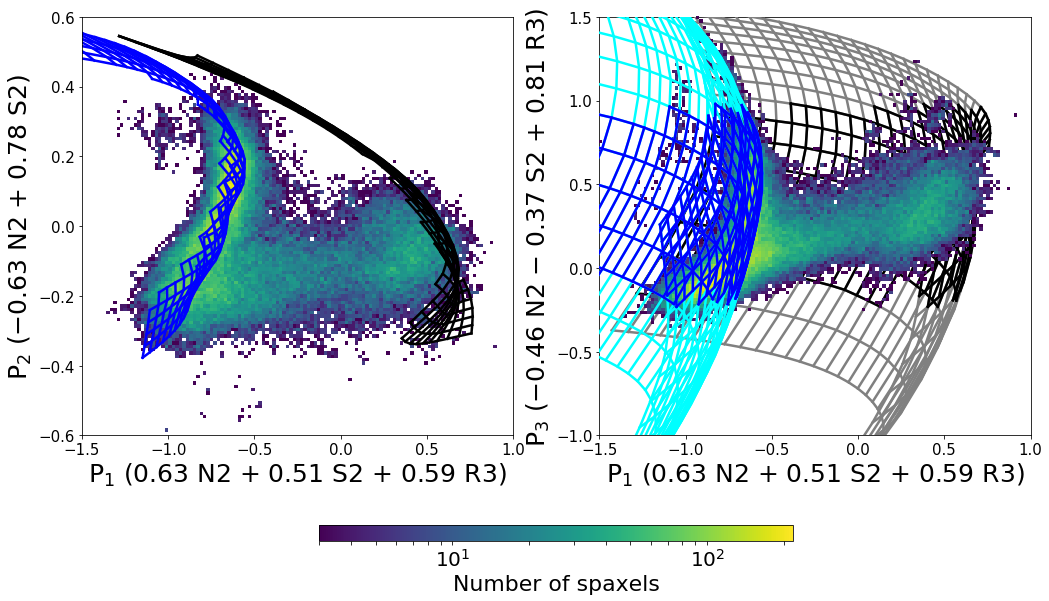}
	\caption{Left panel: final plane of projection. The photoionzation models are first interpolated and then cut in the $P_3$ (perpendicular) direction so that they cover the middle 98\% of the data points. Right panel: $P_1-P_3$ plane. Parts of the model grids (in cyan and grey) are cut off according to the distribution of the data points along the $P_3$ axis. The 2D number density distributions of the our sample spaxels are displayed in both panels.}
    \label{fig:newr_2d}
\end{figure*}

With the rotation angles determined, we calculate the expressions for the axes of this reprojected plane, which are:
\begin{equation}
    P_1 = 0.63 \text{ N2} + 0.51 \text{ S2} + 0.59 \text{ R3} ,
\end{equation}
and
\begin{equation}
    P_2 = - 0.63 \text{ N2} + 0.78 \text{ S2} .
\end{equation}
In the left panel of Fig.~\ref{fig:newr_2d}, we plot our data and models in this new diagnostic diagram. We cut the model grids to keep only those models that cover a similar range on the $P_3$ axis (which is perpendicular to the $P_1-P_2$ plane in the 3D space) as our data. The information from this third dimension further shrinks the spread of the models in the $P_1-P_2$ plane. The expression for $P_3$ is:
\begin{equation}
    P_3 = -0.46 \text{ N2} - 0.37 \text{ S2} + 0.81 \text{ R3} .
\end{equation}
We note that one has the freedom to make a translation to the origin along the $P_3$ axis, which is equivalent to adding a constant term to the above equation. The sample distribution on the $P_1-P_3$ plane is shown in the right panel of Fig.~\ref{fig:newr_2d}. 
98\% of our data lie in the range $-0.23 < P_3 < 0.98$. Thus, we keep only the parts of the model grids with $P_3$ within this range.

The final photoionization models appear quite compact in this $P_1-P_2$ projection. This leads to strong constraints on the model parameters, as the data points for pure AGN-ionized or SF-ionized clouds should trace the model surfaces. The data show a curved locus on the left side, with $P_1$ between $-1$ and $-0.5$, which is traced nicely by our fiducial star-forming model. To the right of it, we see a nearly horizontal extension, with a concentration of points around $P_1\sim0.5$. This blob on the right contains the Seyferts and LI(N)ERs identified by the original BPT diagram. Our fiducial AGN model roughly traces the envelope of this blob rather than going through the densest part of it. This could mean that our fiducial model is not a good description for a typical AGN-photoionized cloud. We will investigate this in the next section. The stretch of points between the SF locus and the AGN blob are far away from both models. As we will demonstrate in the next section, they cannot be explained by varying the parameters of the SF model or the AGN model. The most natural explanation is that they are composites of SF and AGN, due to insufficient spatial resolution of our observations or spatial projection effects in the target galaxy.

Another interesting fact of this new projections is that the coefficients of N2 and S2 for the $P_2$ axis are very close to be opposites, making $P_2$ close to the S2N2 ratio, which is a good tracer for metallicity (\citealp{2016Ap&SS.361...61D}). Thus, we expect the constant metallicity lines in this diagram to have roughly constant $P_2$ values. On the other hand, for the part of the model grids we kept, the constant ionization parameter lines have roughly constant $P_3$ values, as is shown in the right panel of Fig.~\ref{fig:newr_2d}. The outcome of these facts combined is that each iso-metallicity surface almost appears as a line parallel to the $P_1$ axis in the regime where our data points are populated, which facilitates the decomposition of the different ionizing components for our sample, as we will discuss in the next section.

We list in the following several potential applications of this new diagram.
\renewcommand{\labelenumi}{\arabic{enumi}}
\begin{enumerate}
    \item Separate the pure SF regions, pure AGN regions, and composite regions with no ambiguity. Since this projection already implicitly enforces consistency between the [N~II] and [S~II] BPT diagrams, and significantly reduces the projected areas of theoretical model grids, data points of pure SF and pure AGN should distribute about the projected models, and those of the composite regions should lie in the gap between them. One potential uncertainty is that there could be variations of the model positions due to different sets of model parameters. This point is further investigated in \S\ref{subsec:mparam}. The demarcations for different ionized regions in the new diagram based on the decomposition of line ratios are given in Section \S\ref{subsec:decompose}. 
    \item Constrain the photoionization model parameters, like the secondary nitrogen abundance and the shape of the ionizing SED. The data distribution span a narrow range, less than 0.8 dex, along the $P_2$ axis, which is very sensitive to the amount of nitrogen included. Changing the nitrogen prescription should result in noticeable shift of the model relative to the data in the vertical direction, allowing one to put strong constraints on the nitrogen prescriptions. On the other hand, changing the hardness of the input SED will systematically lower/increase all line ratios. Since the vertical axis tends to cancel out the changes in line ratios, the model will move in the horizontal direction.
    \item Estimate metallicity for composite regions. 
    For the composite regions with line ratios sitting between pure-AGN and pure-SF model surfaces, it is potentially possible to construct iso-metallicity lines between the two edge-on models and calculate the metallicity for each point. However, there is still certain difficulty due to the nitrogen prescription inconsistency between the best-fit AGN and SF models. We will illustrate this in \S\ref{subsec:metal}.
    \item Calculate the contributions of different ionizing sources given the corresponding models. Since regions ionized by a single source should ideally have a very narrow distribution about the corresponding model surface, we are able to set up a parameterization so that every point lying between two different ionizing models can be described by the relative contributions from these sources. These results can be used to create demarcation lines with a specific threshold on the contamination from the other sources. But one should be cautious about the assumptions made before decomposing the sources.
\end{enumerate}
We will give some preliminary examples of these applications in the next section.

\section{Implications and Applications of the new diagram}
\label{sec:application}

In this section we discuss both the implications and applications of the new projection for the optical BPT diagnostics.

\subsection{Constraints on model parameters}
\label{subsec:mparam}

With our new $P_1-P_2$ projection, we consider that the correct SF model should go through the densest part of the SF locus on the left and the correct AGN model should go through the densest part of the AGN blob on the right. There are two reasons. First, the data should show scatters around the model due to intrinsic scatter of the physical parameters and observational uncertainty on the line ratios. Second, for the composite objects, the most natural line ratio distribution of them should pile up at the two end points where one component completely dominates over the other. Any concentration of data in-between the end points would require an explanation of why that particular ratio of combination is preferred. Therefore, we think the densest regions of the SF locus and AGN blob should match the pure SF and AGN models, respectively.


Thus, we first examine how the parameters of the photoionization model can impact the position of the models in this new diagram. In particular, we need to find what changes to the AGN model parameter could lead to a better fit to the densest part of the AGN blob. We will see that the data distribution in this new projection can place stringent constraints on the models. We consider three important inputs to the photoionization models: the hydrogen density of the cloud, the secondary nitrogen abundance (since we are using the [N II] emission line), and the shape of the input ionizing SED.

\begin{figure}
	\includegraphics[width=0.5\textwidth]{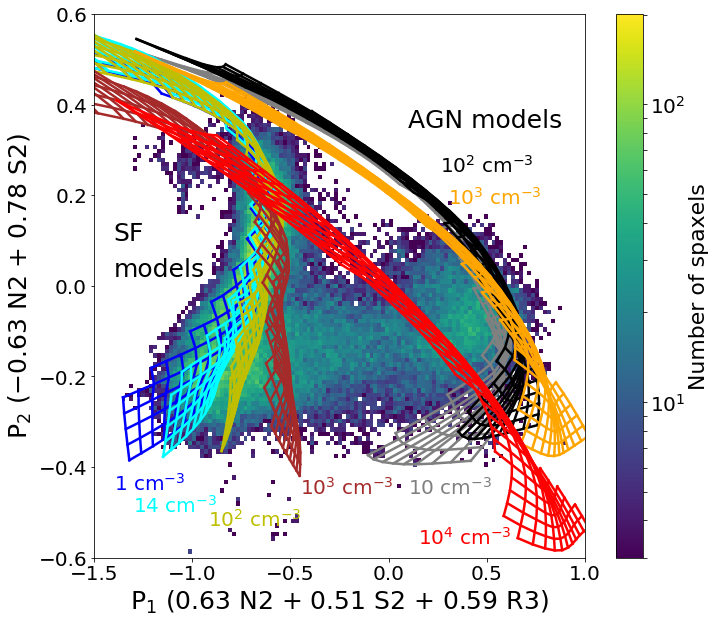}
	\caption{SF and AGN models with various hydrogen densities in our new projected diagram. Our fiducial SF model has a density of 14 cm$^{-3}$, and our fiducial AGN model has a density of 100 cm$^{-3}$. For SF models, we compute four densities from 1 cm$^{-3}$ to $10^3$ cm$^{-3}$, and for AGN models, we compute four densities from 10 cm$^{-3}$ to $10^4$ cm$^{-3}$. The densites values are shown by the side of the corresponding models. All the models are cut so that they enclose the middle 98\% of the data along the $P_3$ axis.}
    \label{fig:newr_den}
\end{figure}

Among the three inputs, the hydrogen density is well constrained by the electron density derived using the [S~II]$\lambda$6716/[S~II]$\lambda$6731 ratio of our sample. The {\tt PyRAF} calculation gives a median value of 100 cm$^{-3}$ for the AGN spaxels and 14 ${\rm cm}^{-3}$ for the SF spaxels in MaNGA. This is in good agreement with the result we obtain by directly comparing the [S~II] doublet ratios produced by {\tt CLOUDY} models to the median ratio in the data. 
Thus, we do not need to change this assumption. Regardless, we illustrate here how density variation will impact the models in our new projection, which demonstrates the power of looking from the right angle. Fig.~\ref{fig:newr_den} shows this density dependence of the models. At low metallicities and when density is less than 100 ${\rm cm}^{-3}$ for SF or 1000 ${\rm cm}^{-3}$ for AGN, the models are not sensitive to the densities. But at high metallicities, both the SF and AGN models shift to the right with an increasing density. At even higher density, the AGN model grid moves downward (at high metallicity) and even to the lower left (at low metallicity). The movement of the model surface is mostly due to the temperature dependence of the collisional excitation rates for [S~II] and [N~II]. The change in direction above a density of 1000 ${\rm cm}^{-3}$ is due to the collisional de-excitation of [S~II]. These movements are very significant compared to the data concentration. In the traditional BPT diagrams (see Fig. 11 of \citealt{Ji2020} for example), the role of density is much less obvious because it is all hidden by projection effects due to the non-optimal viewing angles. 

\begin{figure*}
	\includegraphics[width=1.\textwidth]{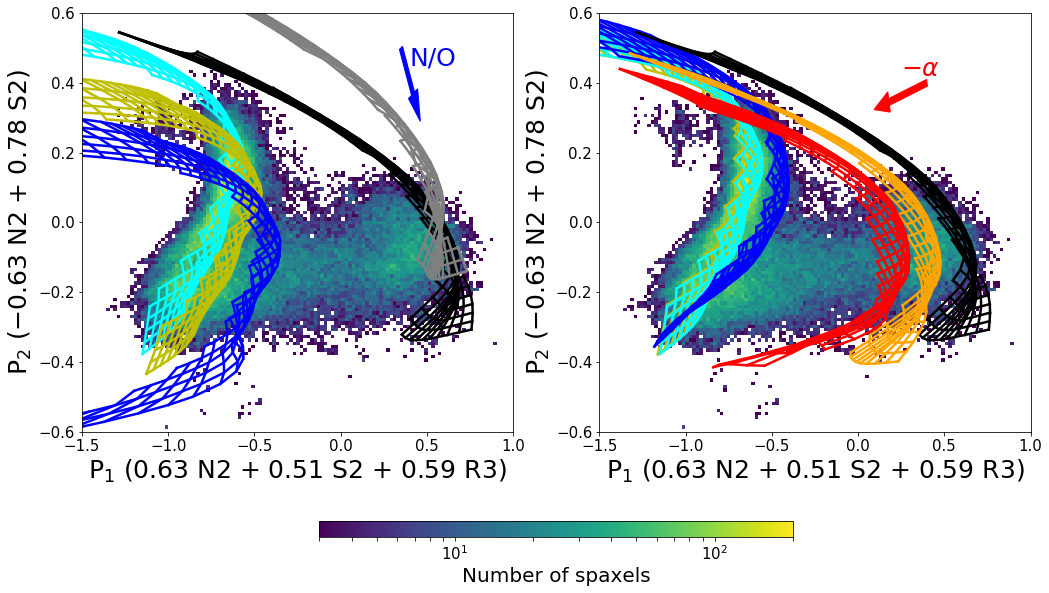}
	\caption{Photoionization model grids and the distribution of MaNGA central spaxels in the new diagram. The color-coding of the data indicates the density of spaxels in the diagram. Left panel: the SF model with the Dopita13 (Groves04) nitrogen prescription is shown in cyan (blue), and the AGN model with the Dopita13 (Groves04) nitrogen prescription is shown in grey (black). Besides these two extreme prescriptions, the Scheafer20 prescription is used to compute the yellow model grid for SF regions. The blue arrow roughly indicates how the model grids would move if the overall N/O ratio is increased. Right panel: AGN models with a Groves04 nitrogen prescription and input SEDs of which the power-law indices are $-1.4$, $-1.7$ and $-2.0$ are shown in black, orange, and red, respectively. The red arrow roughly indicate in what direction the AGN model grid would move if the power-law index of the input SED is decreased. SF models generated by SEDs of continuous star-formation history but different ages are plotted as well. The star-forming ages for the blue, cyan, and yellow models are 0.01 Myr, 4 Myr, and 8 Myr, respectively. All the models are cut so that they enclose the middle 98\% of the data along the $P_3$ axis.}
    \label{fig:newr_mp}
\end{figure*}

\begin{figure}
	\includegraphics[width=0.5\textwidth]{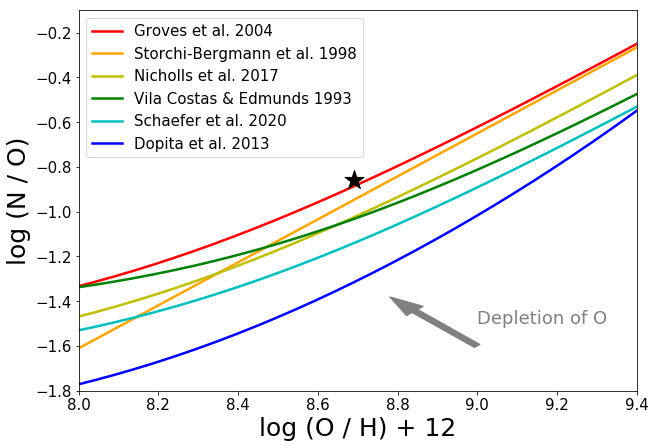}
	\caption{Various nitrogen prescriptions in literature. The Groves04 and the Dopita13 are the two prescriptions used in this paper. The solar abundance is indicated by the black star. The arrow shows the effect of the depletion of oxygen onto dust grains by 0.22 dex. By default {\tt CLOUY} assumes that oxygen depletes by 40\% or approximately 0.22 dex, while nitrogen does not deplete.}
    \label{fig:np_com}
\end{figure}

In comparison to the density, the nitrogen abundance prescription is more difficult to measure, which results in significant disagreements in the literature. Though \cite{2020ApJ...890L...3S} derived a median N/O ratio vs. metallicity relation for H~II regions from MaNGA, they also find that this relation depends on the total stellar mass and the local star formation efficiency of the host galaxies. It is possible that AGN-ionized regions (and/or LIERs) have different nitrogen enhancement from that of the SF regions, as they are more likely to be found in more massive and evolved galaxies. In order to constrain this parameter from the side of theoretical models, we need to understand how the model grids will move as a whole when the nitrogen prescription is changed. Qualitatively speaking, as the nitrogen abundance at a given metallicity increases, the N2 ratio should increase as well. This makes the model grid move in the lower right direction, as indicated in the left panel of Fig.~\ref{fig:newr_mp}. The exact behavior of the model grid might well differ for parts with different metallicities, as can be seen in the case of the AGN models, where the model also seems to rotate a little bit after we switch its nitrogen prescription. The two sets of AGN models we plot in this figure are generated with nitrogen prescriptions adopted by \cite{2004ApJS..153....9G} and \cite{2013ApJS..208...10D} (we denote them as Dopita13 prescription and Groves04 prescription hereafter), respectively. The corresponding log(N/O) vs. 12 + log(O/H) relations are shown in Fig.~\ref{fig:np_com}, together with some other relations we found in the literature \citep[cf.][]{1993MNRAS.265..199V, 1998AJ....115..909S, 2017MNRAS.466.4403N, 2020ApJ...890L...3S}. We choose these two relations because they yield the highest and the lowest N/O ratios for all metallicities. The separation between the two sets of model grids basically shows the largest difference in line ratios induced by the variation in the N/O ratio. For SF regions, we add another model with the nitrogen prescription derived by \cite{2020ApJ...890L...3S} (hereafter Schaefer20 prescription), which gives the second lowest N/O ratios for most metallicities as shown in Fig.~\ref{fig:np_com}. To fit the SF locus traced by the densest part of the data distribution, the lowest N/O ratio prescribed by \cite{2013ApJS..208...10D} is needed. A slightly elevated N/O vs. [O/H] relation like the Schaefer20 prescription would be enough to drive the model away from the center of the SF locus. As a consequence, the SF locus in this diagram puts a strong constraint on the nitrogen prescription for our SF models. On the other hand, an AGN model with the Dopita13 prescription would lie too high to match the AGN blob in our new diagram centered around $P_1\sim0.4$ and $P_2\sim-0.1$.

As a result, it seems that there is a discrepancy between the best fitting nitrogen prescriptions indicated by the SF and AGN models. Raising the hydrogen density of the AGN model to much higher than $1000~{\rm cm}^{-3}$ could lower its vertical position. But this is not a viable solution as it would contradict the measured [S II] doublet ratio.

Another model input we can vary is the SED of the ionizing spectra. The right panel of Fig.~\ref{fig:newr_mp} shows that by softening the input SED of the AGN model, we can move the model grid to the left and slightly downward. An AGN model with a softer SED and the Dopita13 nitrogen prescription could match the upper part of the AGN blob, but would require unrealistically high metallicity ([O/H] $>0.75$) to cover the bottom part of it. Therefore, varying the SED would not solve the discrepancy in the nitrogen prescription between SF and AGN models. 

In fact, if we use the density distribution of the AGN blob as a metric, to make the AGN model go through the densest part of it requires a Groves04 nitrogen prescription and a power-law index of $\sim ~-1.7$, as shown in the right panel of Fig.~\ref{fig:newr_mp}. 

If we consider adjusting the nitrogen prescription of the SF model instead, the SF model with a Groves04 nitrogen prescription would fall too far to the lower right of the SF locus. It seems even harder to move it up in the upper left direction by changing the density or the shape of the ionizing SED. In the right panel of Fig.~\ref{fig:newr_mp}, we show several SF models with different input SEDs corresponding to different ages for a starburst with constant star formation rate. The models do not differ in their vertical positions at all. According to \cite{2019ApJ...878....2D}, possible variations in the ionizing radiation field of the H~II regions from different population synthesis models only introduce an average systematic difference of $\sim$ 0.1 dex in the N2 and R3 line ratios. Thus, varying the input SEDs for the SF models also could not resolve this discrepancy.

To summarize, solely judging from the distribution of the data in our new diagram, the best-fit SF model has a Dopita13 nitrogen prescription and a continuous SFH no younger than 4 Myr, while the best-fit AGN model has a Groves04 nitrogen prescription and a power-law index of $-1.7$. This discrepancy in the nitrogen prescription cannot be easily solved by adjusting other model parameters. This discrepancy suggests some assumptions we make in modelling the ionized cloud must be incorrect, or there are some difference in the physical environments around SF and AGN that we have not yet considered in our modeling. We will further discuss this puzzle and its consequence in the next section.


\subsection{AGN/SF composites and the iso-metallicity mixing sequence}
\label{subsec:metal}

With the above exploration of how AGN and SF models depend on the model parameters and assumptions, we found that none of those models could completely cover the zone between the SF locus and the AGN blob. Thus, the simplest explanation is that these data points are from truly composite regions made up by combining SF-ionized clouds with AGN/LIER-ionized clouds.

We would like to clarify what we consider as the correct physical picture for a composite region. We consider that a composite region is composed of spatially close but physically separate clouds with different ionizing sources, some are ionized by massive young stars (SF), and others by AGN or whatever source is responsible for LIERs or diffuse ionized gas. The spectra of these clouds are combined together due to insufficient spatial resolution of the observations or projection effects in the target galaxy, leading to a set of line ratios that is intermediate between SF and AGN/LIERs.

We consider that it is unlikely for the spectra of composite regions to come from clouds that are ionized by a composite SED made up by comparable fractions of AGN photons and SF photons, as it requires fine tuning to make them to be on the same order of magnitude. Given that radiation flux falls with distance squared, it is more likely that they differ by orders of magnitude for a random location in the galaxy. Thus, a single cloud is nearly always dominated by one of the ionizing sources. In galaxies with significant shock ionization, the same principle should apply that each cloud is dominated by one of the ionization mechanisms rather than balanced by comparable contributions from more than one.

Under this physical picture, it is reasonable to assume the AGN/LIER-ionized clouds and SF-ionized clouds that make up a composite region are spatially close (on scales smaller than the PSF of the observations) and have roughly the same metallicities and abundance pattern.

Given this picture and this assumption, it is now feasible to decompose the composite regions into AGN and SF components, and at the same time, constrain their metallicities. Our new projection of the line-ratio space makes these constraints quite apparent and much less ambiguous than decompositions based on the original [N~II]- and [S~II]-based diagrams. As long as we can find consistent abundance prescriptions between AGN and SF regions, we can connect those AGN models with SF models at the same metallicity and define iso-metallicity mixing sequences. The fact that the ionization parameter is nearly hidden in this $P_1-P_2$ projection means the decomposition is nearly independent of the ionization parameters of the AGN and SF regions.

However, as we discussed in \S\ref{subsec:mparam}, we cannot find a single nitrogen prescription to fit both the SF locus and the AGN blob. Without such a consistent abundance pattern prescription, we cannot carry out a self-consistent decomposition and metallicity determination for composites. Regardless, we show here how the iso-metallicity sequences would look like in our new projection, and how they compare to the distribution of composite regions, which will further demonstrate the tension in the nitrogen prescription between the SF and AGN models.

We begin by setting up a series of parameterized iso-metallicity surfaces (or line bundles) connecting our best fitting AGN model and SF model in the 3D diagram. Since we adopt different nitrogen prescriptions for the AGN and SF models, the `iso-metallicity' here really means iso-[O/H], but not iso-(N/O). We show below an example of how we parameterize a mixing line between a pre-determined start point on the AGN surface and an end point on the SF surface. The start point is assumed to have line ratios of a pure AGN region, which we denote as $\rm N2_{AGN}$, $\rm S2_{AGN}$, and $\rm R3_{AGN}$. Similarly, the end point corresponds to a pure SF region with line ratios $\rm N2_{SF}$, $\rm S2_{SF}$, and $\rm R3_{SF}$. Now if we observe the two regions together, the resulting line ratios will change, according to the relative strength of the emission lines from different regions. Take N2 as an example, the observed value $\rm N2_{obs}$ can be calculated using the following equation.
\begin{equation}
    \rm N2_{obs} = log (\frac{([N~II]/H\alpha )_{AGN} + {\it i}~ ([N~II]/H\alpha )_{SF}}{1+\it i}),
\end{equation}
where $i$ is defined as the ratio of the H$\alpha$ flux contributed by the SF region to that contributed by the AGN region, i.e. $i\equiv \rm H\alpha _{SF}/H\alpha _{AGN}$. $\rm S2_{obs}$ can be computed using a similar equation. Whereas the equation of $\rm R3_{obs}$ involves the SF-to-AGN ratio for H$\beta$ flux, which we denote as $j$. Note $j$ is not necessarily the same as $i$ because the SF-ionized clouds and the AGN-ionized clouds could have very different dust extinction, although they are spatially close. To relate $i$ and $j$, we need to assume a ratio between the Balmer decrement for the AGN clouds ($b\rm _{AGN}$) and that for the SF clouds ($b\rm _{SF}$), which we denote as $\Theta \equiv b_{\rm AGN}/b_{\rm SF}=j/i$. We can make a simplification by choosing a constant value for $\Theta$. By taking the ratio of the median Balmer decrement among the AGN regions to that among the H~II regions in our sample, we obtain a representative value of 1.35 for $\Theta$. We will use this value in the following analysis, i.e. $j=1.35i$. Difference in $\Theta$ would lead to a different decomposition result but will not shift the iso-metallicity sequence significantly.

As a result, there is effectively one parameter for constructing a single mixing sequence that connects a start model point and an end model point. We can parametrize this sequence by changing log($i$) from $-\infty$ to $\infty$. By varying the ionization parameter of the start model point and the end model point at fixed metallicity, we can generate a bundle of such lines describing the overall trend of an iso-metallicity mixing sequence.

\begin{figure}
	\includegraphics[width=0.5\textwidth]{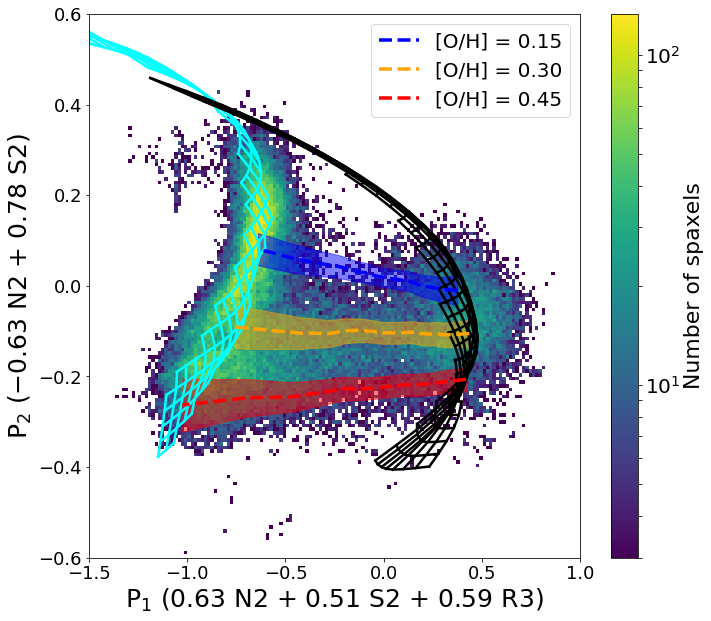}
	\caption{Examples of iso-metallicity line bundles constructed from the photoionization models. The dashed lines are the median trends of line bundles with different metallicities. The colored shaded regions represent the projected areas within one stander deviation from the median along the vertical axis.}
    \label{fig:iso_zlines}
\end{figure}

Fig.~\ref{fig:iso_zlines} shows three examples of the projected iso-metallicity surfaces (or line bundles) with [O/H] = 0.15, 0.30, and 0.45. The spread of each line bundle is caused by the variation in the ionization parameter. To fully utilize the information contained in the 3D distribution of the data, we re-cut the model grids so that they only enclose the middle 98\% composite spaxels (i.e. excluding the upper part of the SF locus with roughly $P_3>0.4$ as shown in the right panel of Fig.~\ref{fig:newr_2d}). While the median- and high-metallicity line bundles appear nearly horizontal, the low-metallicity line bundle show a slightly negative slope. It appears that the composite spaxels in MaNGA  follow a similar trend as our theoretical iso-[O/H] sequences. However, we note that the two ends of the composite sequence in the data are not necessarily having the same metallicity as they could come from different galaxies.

\begin{figure*}
	\includegraphics[width=1.\textwidth]{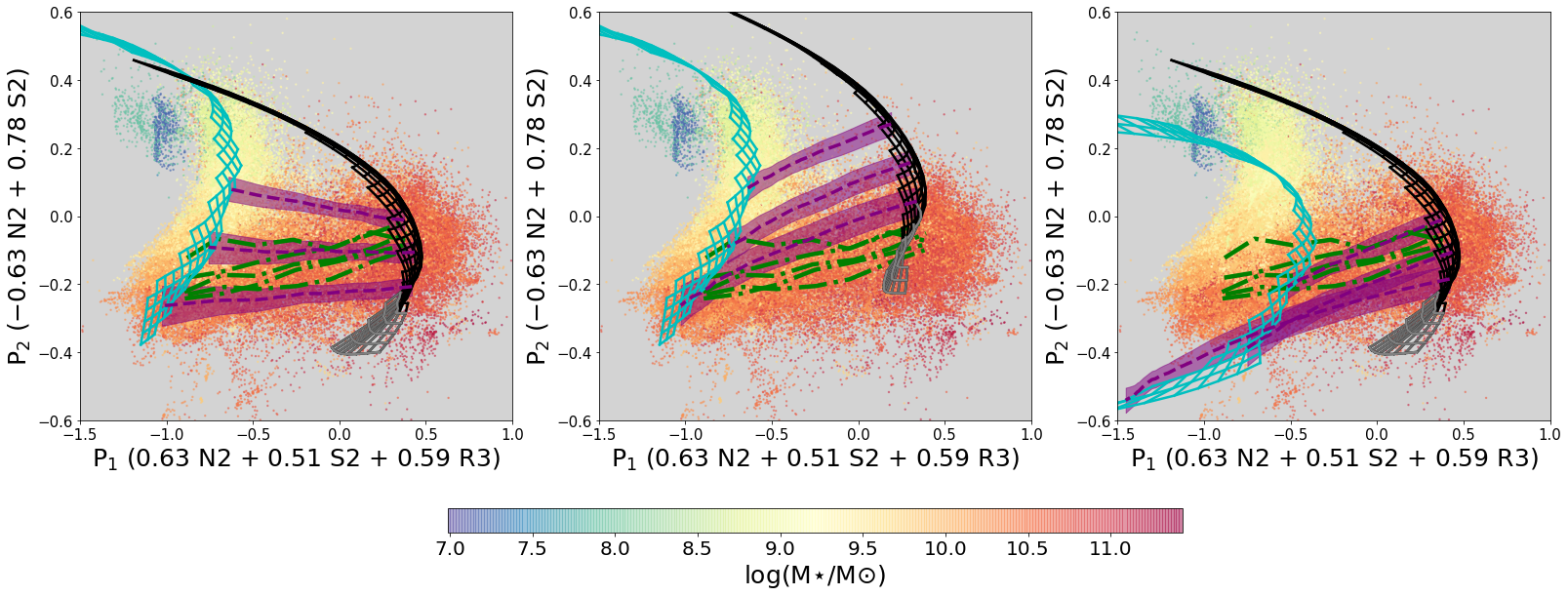}
	\caption{Comparisons of iso-metallicity line bundles based on different combinations of SF (cyan) and AGN (black and grey) models. The three purple bands correspond to three iso-metallicity line bundles in with [O/H] = 0.15, 0.30, and 0.45. All spaxels are color-coded according to the total stellar masses of their host galaxies. The dotted-dashed green lines trace the median trends of five stellar mass bins, from $\log \rm (M_{\star}/M_{\odot})=9.75$ to $\log \rm (M_{\star}/M_{\odot})=11.00$ with an interval of 0.25 dex. In the left panel, a SF model with the Dopita13 nitrogen prescription and an AGN model with the Groves04 nitrogen prescription are shown; in the middle panel, both models are generated using the Dopita13 prescription; in the right panel, both models are generated using the Groves04 prescription. The parts of the AGN models with [O/H] > 0.5 are colored in grey.}
    \label{fig:com_zlines}
\end{figure*}

A sanity check for whether our models produce a sensible composite sequence is to compare the derived iso-metallicity lines with the overall shape of the composite zone. In Fig.~\ref{fig:com_zlines}, we perform a more careful comparison by binning our data according to the total stellar masses of their host galaxies. This would help identify iso-metallicity composite sequences in the data by making sure the two ends of them are from similar types of galaxies. The stellar masses are drawn from the MaNGA Firefly value-added catalog \citep{2017MNRAS.466.4731G}. Within each stellar mass bin, we expect both [O/H] and N/O are roughly fixed with small scatters, and the distribution of the spaxels in the composite zone provides the correct slope for an iso-metallicity line bundle. Here we are assuming that the mass-metallicity relations for SF galaxies and AGN host galaxies are identical (for a potential counterexample, however, see \citealp{Thomas19}). The theoretical iso-[O/H] line bundles produced by our best-fit models all exhibit shallower slopes than those indicated by the median data trends in five different mass bins, which include most spaxels in the composite zone. To steepen the slopes of these iso-metallicity lines by changing the abundance pattern of the photoionization models, one could either lower the N/O vs. [O/H] relation for the AGN model, or elevate this relation for the SF model. In the middle and right panels of Fig.~\ref{fig:com_zlines}, we plot the iso-metallicity lines using SF and AGN models with consistent nitrogen prescriptions, Dopita 13 for the middle panel and Groves04 for the right panel. We can see that lowering the N/O ratio raises the position of the AGN model in the diagram, and the resulting slopes of the iso-metallicity lines are now steeper compared with the data. A noticeable problem with lowering the N/O ratio of the AGN model is that the composite zone cannot be fully covered. Since the SF locus only reaches a highest [O/H] of 0.5 according to our SF model, if we only consider the corresponding part of the AGN model (which is colored in black), even the iso-metallicity line bundle with the highest [O/H] could not overlap with the lower part of the mixing branch. On the other hand, if we raise the N/O ratio of the SF model instead, as is shown in the right panel of Fig.~\ref{fig:com_zlines}, the slopes indicated by our models for the composite zone are again too steep, and the SF model does not match the SF locus in the data.

To summarize, the nitrogen prescription for the SF model is strongly constrained by this SF locus, as we have already seen in \S\ref{subsec:mparam}. However, in order to cover the AGN blob in the data, the AGN model needs a prescription with higher N/O ratios, which results in an inconsistency in the abundance pattern to explain the composite galaxies. Furthermore, forcing the SF and AGN models to have the same N/O prescriptions makes the iso-metallicity mixing-sequences to be too steep compared to the trends of composite galaxies with fixed stellar masses. It could be due to incorrect assumptions or missing physics in our photoionization models, and we will investigate this issue with more sophisticated photoionization models in future work. Regardless, this illustrates nicely the power of our new projection of the line ratio space.

\subsection{Decomposition of the ionized regions}
\label{subsec:decompose}

In the previous section we introduce a way to construct iso-metallicity line bundles (or surfaces) between the AGN and SF model grids, assuming the Balmer decrement ratio, $\Theta$, is fixed. As a result, there is effectively one free parameter that determines the position of a point on a given iso-metallicity line, which we can choose to be the ratio of the H$\alpha$ flux from the SF-ionized clouds to that from the AGN-ionized clouds, $i$. Alternatively, we can use $i$ to derive another physical parametrization. The fractional contribution to the total H$\alpha$ flux by the AGN-ionized clouds is H$\alpha \rm_{AGN}$ / H$\alpha \rm_{total}$ = $1/(1+i)$, which we denote as $f\rm _{AGN}$ hereafter. This parameter tells us how significant the contribution from AGN-ionized clouds is in a composite spectrum.

The decomposition process is summarized as the following. First, using a pair of pre-determined SF and AGN models, we construct all possible iso-metallicity lines (for all ionization parameter combinations), with the $i$ parameter ranging from $10^{-3}$ to $10^3$ equally spaced in the logarithmic space. For each data point, we compute the probability distribution of the models using Bayesian statistics, assuming a flat prior (i.e. equal prior probability for all model grid points). We compute the likelihood of each data spaxel given each model point in the original 3D parameter space (N2, S2, R3) with the uncertainty in these three line ratios (in log). We do the calculation in the original 3D space as there is less covariance among them than there is among $P_1$, $P_2$, and $P_3$. Given the resulting probability distribution of the models, we compute the expected value for $\log i$ and convert it to $f_{\rm AGN}$ for each spaxel.

The choice of photoionization models will have a non-negligible influence on the decomposition result. Since we currently cannot settle the nitrogen discrepancy while matching the models to the data, there should be no self-consistent way to decompose the data either. Still we would like to show in the following a demonstration of the decomposition using our best-fit SF and AGN models (the set shown in the left panel of Fig.~\ref{fig:com_zlines}). Given that these models can well describe the majority of the AGN and SF spaxels, they should not be far from the correct models. But the correct models could have different slopes for the iso-metallicity lines. Judging from the left panel of Fig.~\ref{fig:com_zlines}, the bias on $f_{\rm AGN}$ due to different slopes of the iso-metallicity lines should not be too significant.

\begin{figure}
	\includegraphics[width=0.5\textwidth]{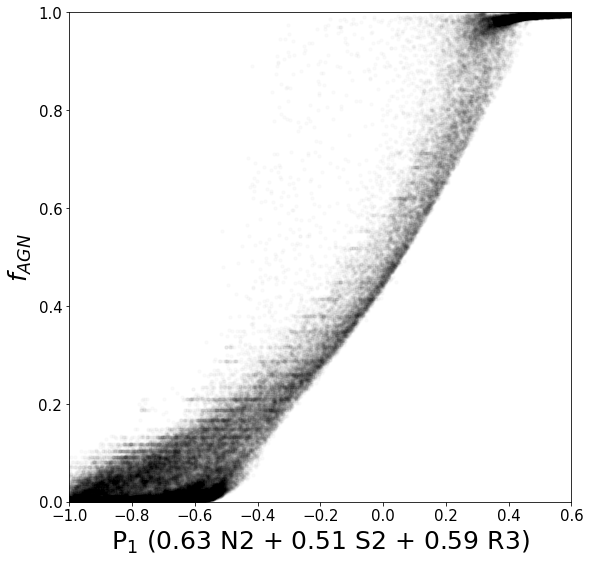}
	\caption{The derived $f\rm_ {AGN}$ as a function of the position of the spaxel along the $P_1$ axis.}
    \label{fig:agnr}
\end{figure}

In Fig.~\ref{fig:agnr}, we plot the derived $f\rm _{AGN}$ as a function of the position along the $P_1$ axis. Due to the spread induced by the variation in the metallicity and the ionization parameter, the relation has relatively large uncertainties at both small $P_1$ values and large $P_1$ values. Part of the data points also seem to cluster at discrete values of $f\rm _{AGN}$, which might be due to our insufficient sampling of log $i$. Despite the spread, we can see that $f\rm _{AGN}$ increases more rapidly as $P_1$ gets larger, but quickly plateaus after $P_1$ reaches 0.3, which is roughly the horizontal position of the lower part of the AGN model. The correlation between $f\rm _{AGN}$ and $P_2$, or that between $f\rm _{AGN}$ and $P_3$, is much weaker.

If the Balmer decrement ratio $\Theta$ is changed, the horizontal position of each parametrized point would change accordingly. Increasing $\Theta$ would shift the model points to smaller $P_1$ values, hence increasing the overall fitted $f\rm _{AGN}$ of our sample. For example, if the $\Theta$ in our setting is doubled (i.e. $\Theta=2.70$ in this case), the fractional change in the derived $f\rm _{AGN}$ of our sample has a median value of 12\%, and the the values for the 16 percentile and the 84 percentile of the data are 2\% and 22\%, respectively. One may wonder if we can use the observed Balmer decrement to put constraints on $\Theta$. This is not sufficient because the observed Balmer decrement also depends on the Balmer decrement in the SF-ionized clouds, $b_{\rm SF}$. Therefore, adding this additional constraint does not remove the need to make one assumption. In principle, if two additional dust-sensitive line ratios were added as extra constraints, it would be possible to solve for both $b_{\rm SF}$ and $\Theta$.

\begin{figure*}
	\includegraphics[width=1.\textwidth]{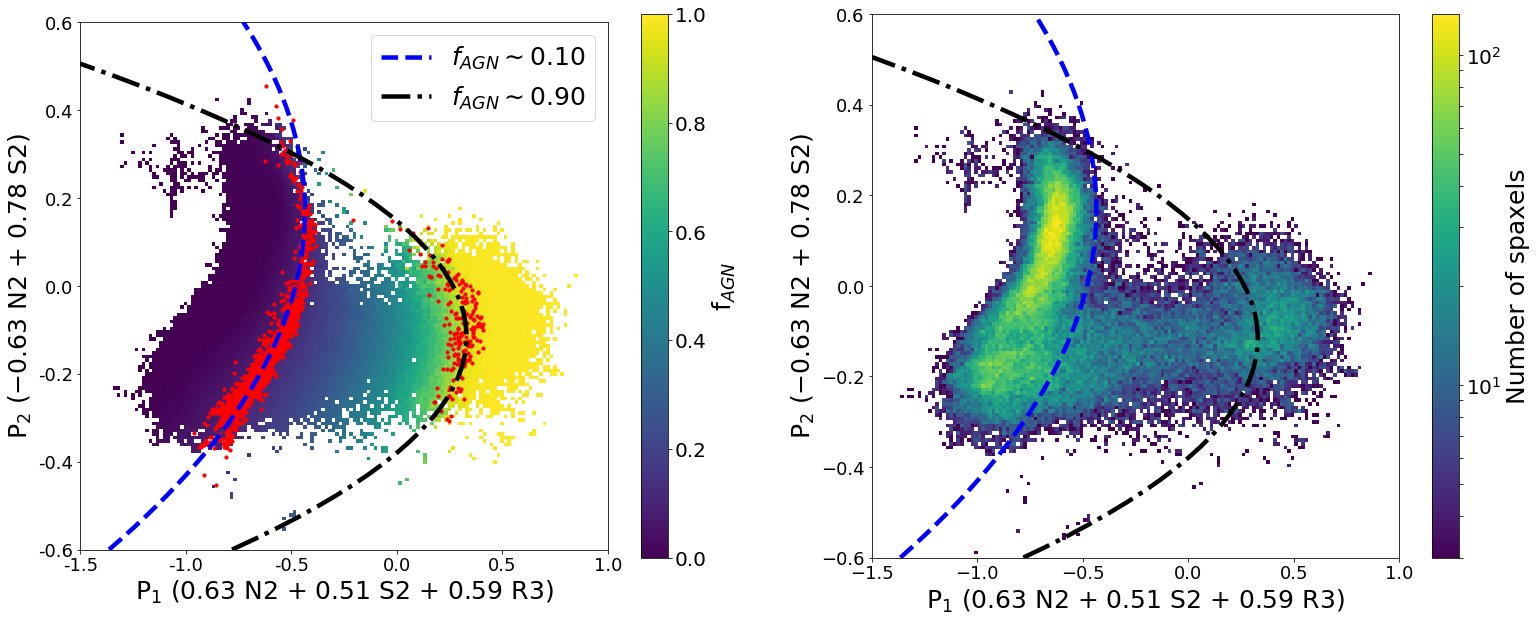}
	\caption{Left panel: map of $f\rm_ {AGN}$ in the new diagram. The dashed blue line and the dotted-dashed black lines are two fitted demarcation lines which correspond to $f\rm_ {AGN}=0.10$ and $f\rm_ {AGN}=0.90$, respectively. The spaxels we select to fit are colored in red, and we use quadratic functions to fit them. Right panel: density distribution of our sample with the same demarcation lines in the new diagram.}
    \label{fig:agnr_map}
\end{figure*}

Fig.~\ref{fig:agnr_map} shows a map of all MaNGA central spaxels color-coded by the average $f\rm _{AGN}$ of each line-ratio bin. With this map, we can generate demarcation lines that correspond to a constant fractional contribution to the total H$\alpha$ flux by AGN-ionized clouds. The dashed blue line corresponds to $f\rm _{AGN}\sim 0.10$, which means the spaxels to their left have a maximum contamination of 10\% from AGN-ionized clouds. Similarly, the spaxels to the right of the dotted-dashed black line have a maximum contamination of 10\% from SF-ionized clouds. We can fit $P_1$ as a quadratic function of $P_2$ for these demarcation lines. The functional form of the demarcation line that identifies SF-dominated regions is
\begin{equation}
    P_1<-1.57P_2^2+0.53P_2-0.48~~({\rm for}~ -0.4\lesssim P_2\lesssim 0.1),
\end{equation}
and the functional form of the demarcation line that identifies AGN-dominated regions can be written as
\begin{equation}
    P_1>-4.74P_2^2-1.10P_2+0.27~~({\rm for}~-0.4\lesssim P_2\lesssim 0.1).
\end{equation}
These demarcations are largely determined by the spaxels inside the composite zone, and therefore should only be applied when $-0.4\lesssim P_2\lesssim 0.1$. From the map (and also Fig.\ref{fig:agnr}), we can see that the AGN fraction $f\rm _{AGN}$ drops more rapidly near the start points of the AGN model. The demarcation lines for AGN regions are largely determined by the location of the best-fit AGN model grid. In the right panel of Fig.~\ref{fig:agnr_map}, we compare these two demarcations with the density distribution of our sample. While SF demarcation line clearly traces the outskirt of the SF locus, the AGN demarcation line is close to the center of the AGN blob.

Since we are using the densest area at the tip of the mixing zone to constrain the AGN model, the location might well change for different samples. For example, a sample with systematically higher Eddington ratios might favor a photoionization model generated by an SED with larger optical to X-ray ratio, thus change the locations of our demarcation lines based on the decomposition \citep{2010A&A...512A..34L}. We also note that potential contaminations from LI(N)ERs which have similar line ratios to AGN-ionized clouds but are not truly ionized by AGNs might hamper the determination of the best-fit AGN model. However, if these pure LI(N)ER spaxels cluster at the same location of the true AGN-ionized clouds in the diagram, the decomposition still applies as the positions of the start points remain the same. The only difference is that instead of obtaining $f_{\rm AGN}$, one obtains $f_{\rm LI(N)ER}$ in some cases. The differentiation of Seyferts and LI(N)ERs depends on the $P_3$ dimension. We will further discuss this issue in the following section.

\section{Discussions}
\label{sec:discuss}

\subsection{Presence of LI(N)ERs and their impact}
\label{subsec:liner}

LI(N)ERs are a class of regions that have enhanced low-ionization forbidden lines (e.g. [N~II]$\lambda 6583$, [S~II]$\lambda \lambda 6716,~6731$, and [O~I]$\lambda 6300$) compared to Seyferts, but have weaker high ionization lines (e.g. [O~III]$\lambda 5007$) \citep{1980A&A....87..152H}. Although they are first discovered in the nuclei of galaxies, recent studies show that most galaxies satisfy the line ratio definition for LI(N)ERs have a more extended distribution, and have radial surface brightness profiles and ionization parameter profiles shallower than expected for a compact central source \citep{2006MNRAS.366.1151S, 2010MNRAS.402.2187S, 2012ApJ...747...61Y, 2016MNRAS.461.3111B}. The ionization mechanism for LI(N)ERs is still under debate, possible candidates include low luminosity AGNs, hot evolved stars, shock heating, turbulent mixing layers, and so on \citep[e.g.][]{1993ApJ...407...83S, 1994A&A...292...13B, 1995ApJ...455..468D, 2008ARA&A..46..475H}. It is likely that in a small fraction of objects, a small region around the central supermassive black hole, less than tens of parsecs in size, is truly photoionized by the AGN. But the larger scale emission ($\sim$100 pc and above), which dominates the line luminosity in most non-star-forming galaxies, are unlikely to be photoionized by AGNs, given both the energetics \citep{2010ApJ...711..796E} and spatial profiles \citep{2010MNRAS.402.2187S, 2012ApJ...747...61Y}.

Current empirical classification of LI(N)ERs is based on emission line ratios in the BPT diagrams, where clear bimodal distribution of SDSS galaxies in the mixing branch seems to indicate the existence of two different populations \citep{2006MNRAS.372..961K}. \cite{2011MNRAS.413.1687C} introduce another diagnostic based on [N II]$\lambda$6583 and H$\alpha$ emission lines. Their calculation suggests that regions with the equivalent width of H$\alpha$ line smaller than 3\AA\ are likely to be ionized by hot evolved starts rather than AGNs. More recently, \cite{2019MNRAS.485L..38D} use a diagnostic which combines emission line ratios, radial positions, and velocity dispersions to separation shock ionized regions from AGN regions in their IFU sample.

\begin{figure*}
	\includegraphics[width=1.00\textwidth]{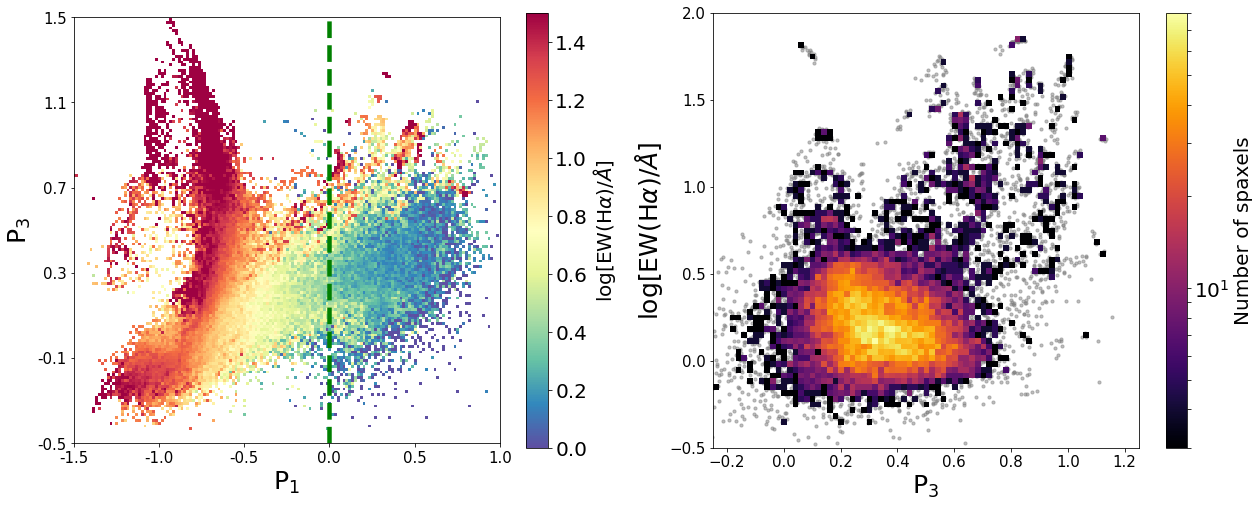}
	\caption{Left panel: map of log[EW(H$\alpha$)] in the $P_1 - P_3$ projection for MaNGA central spaxels. Right panel: log[EW(H$\alpha$)] as a function of the positions along the $P_3$ axis for spaxels with $P_1$ > 0.0, which are to the right of the dashed green line shown in the left panel.}
    \label{fig:ewha_map}
\end{figure*}

\begin{figure}
	\includegraphics[width=0.5\textwidth]{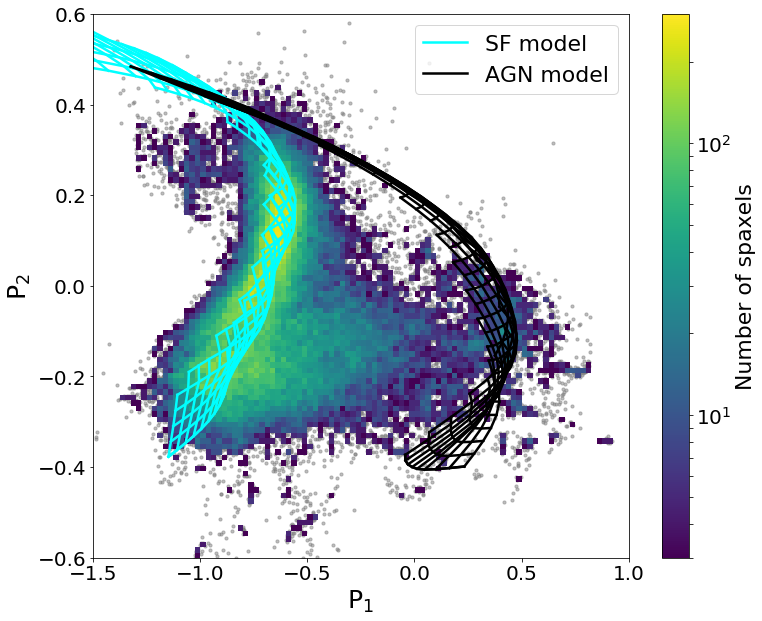}
	\caption{Distribution of MaNGA central spaxels with EW(H$\alpha$) > 3\AA . Although the composite zone is significantly reduced after the cut, the tip of this zone still traces the AGN model grid.}
    \label{fig:3angcut}
\end{figure}

Since LI(N)ERs have relatively low ionization, we expect they are separable from Seyferts in the $P_3$ direction, as it traces the ionization parameter. The left panel of Fig.~\ref{fig:ewha_map} shows the distribution of our MaNGA sample on the $P_1 - P_3$ plane, colored coded by the equivalent width of the H$\alpha$ line (hereafter EW(H$\alpha$)). The EW(H$\alpha$) as a function of the $P_3$ for spaxels with $P_1$ > 0 is plotted in the right panel. Despite that MaNGA has a small fraction of Seyferts, the transition in EW(H$\alpha$) along the $P_3$ dimension is very clear. Viewed on the $P_1 - P_2$ plane, the Seyferts sit on top of LI(N)ERs and the number density at the AGN locus is dominated by LI(N)ERs. If we exclude spaxels with EW(H$\alpha$) < 3\AA\ , as shown in Fig.~\ref{fig:3angcut}, we still find the remaining spaxels cluster around the best fitting AGN model, which ensures that the LI(N)ERs do not bias the location of the densest area of the AGN blob. It also indicates that a large portion of the mixing zone could be a product of mixing between LI(N)ERs and SF-ionzed clouds. The SF-ionized clouds will increase the final EW(H$\alpha$) of the composite spectra, so we expect the 3\AA\ cut to miss more SF-LI(N)ER composite regions near the SF locus. It would be interesting to compare ionization models for non-AGN LIERs in these diagrams with AGN models in future work.

\subsection{SDSS Legacy sample in the new diagram}
\label{subsec:sdss_legacy}

\begin{figure*}
	\includegraphics[width=1.00\textwidth]{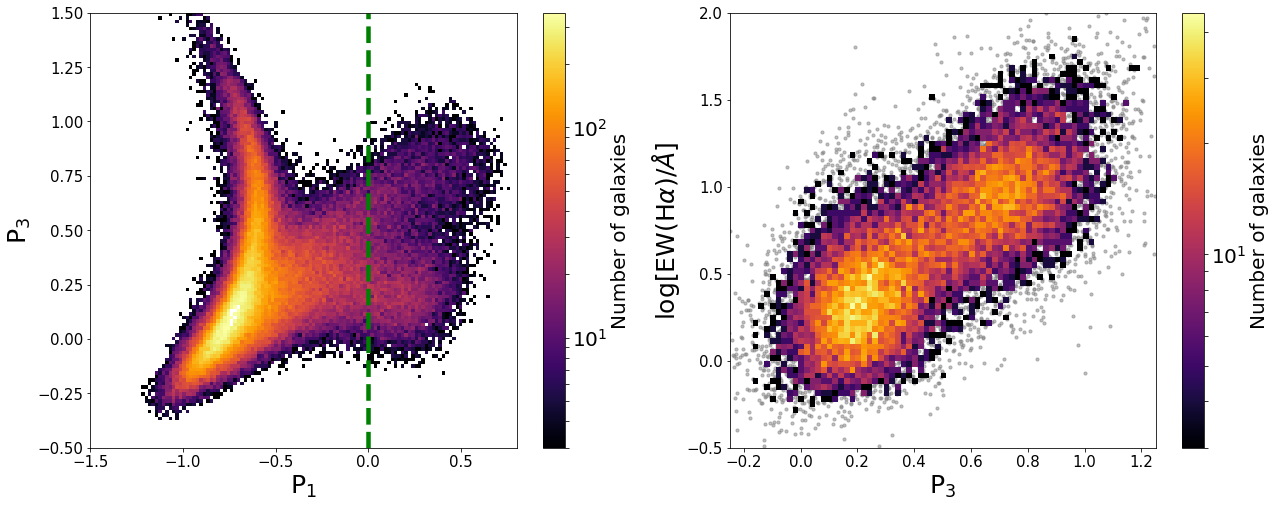}
	\caption{Left panel: density distribution of SDSS galaxies in the $P_1 - P_3$ projection. Right panel: log[EW(H$\alpha$)] as a function of the positions along the $P_3$ axis for galaxies with $P_1$ > 0.0, which are to the right of the dashed green line shown in the left panel.}
    \label{fig:ewha_sdss}
\end{figure*}

The separation of LI(N)ERs and Seyferts would be clearer if we use a sample with a larger Seyfert population. Fig.~\ref{fig:ewha_sdss} shows the original SDSS-I and -II Legacy sample in the $P_1 - P_2$ plane, as well as the correlation between EW(H$\alpha$) and positions of galaxies along the $P_3$ axis for galaxies with $P_1>0$. This sample is drawn from the DR7 of SDSS \citep{2000AJ....120.1579Y, 2009ApJS..182..543A}. The emission line measurements were done using an updated version of the code used by \cite{2006ApJ...648..281Y}, with an additional flux calibration described by \cite{2011AJ....142..153Y}. Zero-point corrections to the equivalent width for emission lines were performed according to \cite{2018MNRAS.481..467Y}, in order to remove systematics from inaccurate stellar continuum subtraction. We select only galaxies with the S/N greater than 3 for all emission lines used in this diagram, as we did for the MaNGA sample. The Seyferts and LI(N)ERs now look comparable in number, and there is an obvious separation between them in the $P_1-P_3$ plane. The transition from LI(N)ERs to Seyferts also seems slightly lower along the $P_3$ axis compared with that in the spatially resolved sample of MaNGA.

For galaxies with $P_1>0$ in this sample, we can see a continuous increase of EW(H$\alpha$) as $P_3$ gets larger. Comparing the right panel of Fig.\ref{fig:ewha_sdss} with that of Fig.~\ref{fig:ewha_map}, we see that the LI(N)ER spaxels in MaNGA with EW(H$\alpha$) < 3\AA\ are able to reach higher $P_3$ values, or equivalently higher ionization parameters, than the SDSS Legacy sample, which is based on single-fiber spectra covering the center 3\arcsec\ of each galaxy. It seems to suggest that spatially resolved LI(N)ERs in MaNGA could have higher ionization. In order to investigate this discrepancy between the two LI(N)ER samples, we select a subsample in MaNGA with $P_1>0.0$, EW(H$\alpha$) < 3\AA\ and $P_3>0.4$, and cross-match it with the SDSS Legacy sample to identify the same galaxies. By selecting MaNGA spaxels within the central 3\arcsec\ aperture and integrating their emission line flux to mimic the SDSS Legacy observations,  we find that the $P_3$ values derived from the integrated MaNGA data is still higher by 0.16 dex than that derived from the SDSS Legacy spectra for the same set of galaxies. Part of this turns out to be due to the lower S/N in the SDSS Legacy data and our selection criteria of requiring S/N greater than 3 in all emission lines. However, even after we remove the S/N cuts for the emission lines, the median $P_3$ values of the two samples still differ by 0.11 dex. We think this remaining offset between the two sets of data for the same galaxies must be due to a combined effect of the difference in the stellar continuum subtraction, the emission-line measurement algorithm, and the zeropoint corrections for the equivalent widths. Using the same data analysis method to process both sample might help to resolve the inconsistency, but this is beyond the scope of this paper and we will leave it to future work.

\subsection{Effect of time evolution on SF models}

Like most studies that use predictions from photoionization models, we assume the modeled ionized cloud to be in a particular evolutionary stage to match the observations. For example, our SF model is based on a stellar SED that corresponds to a continuous SFH of 4 Myr, and an ionized cloud that has swept the surrounding natal cloud and is exposed to the observer. However, as \cite{2020MNRAS.496..339P} suggested, once the effect of time evolution is considered, a typical H II region can have a wide distribution in the [N II] BPT diagram even at a single metallicity. The variations in line ratios associated with these time-dependent models could be potential sources of systematic uncertainties for our calculations, which treat metallicity as one of the primary indicators of the position of a given H II region in the line-ratio space. In the extreme case, zero-age star clusters that are heavily embedded in their natal clouds could produce line ratios similar to those of weak AGNs or DIG due to the internal extinction. However, from the observational point of view, these clouds have very large extinction (with $\rm A_v \gtrsim 3.0$) and cover only a very short period in the evolution history of a typical H II region. Therefore, they are not statistically significant for large samples of H II regions, especially for those with low spatial resolutions, like the MaNGA sample we use (of which the typical $\rm PSF \sim 1.5$ kpc). The SF locus we observed in this work is more likely to be dominated by clouds that are not heavily impacted by differential extinction, as they usually exhibit much higher optical luminosities.

The general effect of this time evolution could still contribute to the scatter we see in the SF locus. But we note that not all evolutionary stages are detectable for extragalactic observations (see Fig.~5 of \citealp{2020MNRAS.496..339P}). Moreover, the data we use in this work are effectively sampling a luminosity-weighted average of different evolutionary stages. Relatively time-steady phases with low internal extinction are more likely to be observed, which is consistent with our model assumptions. Thus we argue that our interpretation of the SF locus being shaped primarily by the variation in metallicity still holds, at least to the first order. Future work that considers the details of the observational effect on selecting the best-fit photoionization models would be useful for studying the time-dependent models.

\section{Conclusions}
\label{sec:conclude}

In this paper we developed a suit of new optical diagnostics different from the original BPT diagnostics, by reprojecting the photoionization models and observational data in a 3D line-ratio space spanned by logarithm of line ratios including N2, S2, and R3. Specific angles for the reprojection are obtained according to the intrinsic shapes and orientations of the SF and AGN photoionization model surfaces in the 3D line-ratio space, so that they both appear nearly edge-on and well separated in one of the new projections. The new diagram clearly reveals the spatial relationships between the data and the models, largely removing projection effects inherent to the original BPT diagrams, and guarantees consistency between data and model in three major line ratios. The applications of the new diagram and the model constraints we obtained can be summarized as the following.

\renewcommand{\labelenumi}{\arabic{enumi}}
\begin{enumerate}
\item The new diagram provides clear classification of spectra among the categories of SF, AGN/LI(N)ER, and Composites with nearly no ambiguity. Demarcations can be defined with a specified contamination fraction from other sources. Composite objects can be decomposed into SF and AGN/LI(N)ERs with well-constrained fractions.
\item With the model surfaces appearing edge-on in our new diagram, the data can put strong constraints on the parameters of photoionization models. We found that the best-fit SF model requires a continuous star formation history not younger than 4 Myr, and the Dopita13 prescription for N/O abundance ratios, while the best-fit AGN model needs a power-law energy index of approximately $-1.7$ between 1-100 Ryd, and an N/O prescription significantly higher than that given by Dopita13. This discrepancy in the N/O prescription between the SF and AGN models cannot be resolved by adjusting other parameters.
\item With both the SF and AGN models well constrained by the data in our new diagram, it is now possible to derive metallicity for AGN/LI(N)ERs and composite objects. However, the discrepancy in N/O prescription between SF and AGN models prevents us from deriving a consistent metallicity solution. Using MaNGA galaxies within narrow stellar mass bins to define iso-metallicity lines and comparing them with the iso-metallicity lines derived from different model prescriptions, we further demonstrated that the discrepancy in N/O prescriptions cannot be avoided in the current set of models.
\item LI(N)ERs occupy the same location as Seyferts in our new $P_1-P_2$ diagram. But they can be clearly separated from each other in the $P_1-P_3$ diagram as $P_3$ traces ionization parameter approximately. The separation corresponds well with the widely-adopted 3\AA\ cut in EW(H$\alpha$), although the exact cut in $P_1-P_3$ plane seems to depend on the details of the data analysis methods.

\item The effect of time evolution of the SF region could potentially contribute to the scatter around the observed SF locus, and introduce systematic uncertainties to our constraints on model parameters. However, the low resolution extragalactic data we use are likely to be biased towards particular evolutionary stages and thus reduce the scatters.

The lesson we learn from this work is that comparing our models with data in a higher dimensional space composed by multiple line ratios can reveal a great deal of information and put strong constraints on the models. Even though higher dimensional space is difficult to visualize, certain projections of it could provide valuable intuitions. Those intuitions could facilitate the construction of algorithms to quantitatively evaluate our models.

\end{enumerate}

\section*{Acknowledgements}

The authors wish to thank David Law for useful comments on a draft of this paper. We acknowledge support by NSF AST-1715898 and NASA grant 80NSSC20K0436 subaward S000353.

Funding for the Sloan Digital Sky Survey IV has been provided by the Alfred P. Sloan Foundation, the U.S. Department of Energy Office of Science, and the Participating Institutions. SDSS acknowledges support and resources from the Center for High-Performance Computing at the University of Utah. The SDSS web site is www.sdss.org.

SDSS is managed by the Astrophysical Research Consortium for the Participating Institutions of the SDSS Collaboration including the Brazilian Participation Group, the Carnegie Institution for Science, Carnegie Mellon University, the Chilean Participation Group, the French Participation Group, Harvard-Smithsonian Center for Astrophysics, Instituto de Astrof\'isica de Canarias, The Johns Hopkins University, Kavli Institute for the Physics and Mathematics of the Universe (IPMU) / University of Tokyo, the Korean Participation Group, Lawrence Berkeley National Laboratory, Leibniz Institut f\"ur Astrophysik Potsdam (AIP), Max-Planck-Institut f\"ur Astronomie (MPIA Heidelberg), Max-Planck-Institut f\"ur Astrophysik (MPA Garching), Max-Planck-Institut f\"ur Extraterrestrische Physik (MPE), National Astronomical Observatories of China, New Mexico State University, New York University, University of Notre Dame, Observat\'orio Nacional / MCTI, The Ohio State University, Pennsylvania State University, Shanghai Astronomical Observatory, United Kingdom Participation Group, Universidad Nacional Aut\'onoma de M\'exico, University of Arizona, University of Colorado Boulder, University of Oxford, University of Portsmouth, University of Utah, University of Virginia, University of Washington, University of Wisconsin, Vanderbilt University, and Yale University.

\section*{Data availability}

The observational data underlying this article are publicly available in the DR15\footnote{www.sdss.org/dr15/data\_access/} and DR7\footnote{classic.sdss.org/dr7/access/} of SDSS (see \S\ref{sec:model_data} and \S\ref{subsec:sdss_legacy}). The derived data generated in this research and data concerning the photoionization models will be shared on reasonable request to the corresponding author.




\bibliographystyle{mnras}
\bibliography{example} 








\bsp	
\label{lastpage}
\end{document}